\documentclass[fleqn,usenatbib]{mnras}
\usepackage{graphics}
\usepackage{graphicx}
\usepackage{color}
\usepackage{aas_macros}
\usepackage{natbib}
\usepackage{amsmath}
\usepackage{comment}
\usepackage{soul} 

\newcommand{\photoz}{photo-\emph{z}}
\newcommand{\Photoz}{Photo-\emph{z}}

\title[PS1-STRM: PS1 classification and \photoz]{PS1-STRM: Neural network source classification and photometric redshift catalogue for PS1 $3\pi$ DR1}

\author[R. Beck et al.]{R\'obert Beck$^{1,2}$\thanks{E-mail: beckrob@ifa.hawaii.edu}, Istv\'an Szapudi$^{1,2}$, Heather Flewelling$^1$, Conrad Holmberg$^{1,3}$, \newauthor 
Eugene Magnier$^1$
\\
$^1$Institute for Astronomy, University of Hawaii, 2680 Woodlawn Drive, Honolulu, HI, 96822, USA \\
$^2$Department of Physics of Complex Systems, E\"{o}tv\"{o}s Lor\'and University, Pf. 32, H-1518 Budapest, Hungary \\
$^3$Platform Services, Stanford Health Care, 300 Pasteur Drive, Stanford, CA, 94305, USA
}

\date{Accepted XXX. Received YYY; in original form ZZZ}

\pubyear{2019}

\begin{document}
\label{firstpage}
\pagerange{\pageref{firstpage}--\pageref{lastpage}}
\maketitle

\begin{abstract}

The Pan-STARRS1 (PS1) $3\pi$ survey is a comprehensive optical imaging survey of three quarters of the sky in the $grizy$ broad-band photometric filters.
We present the methodology used in assembling the source classification and photometric redshift (\photoz) catalogue for PS1 $3\pi$ Data Release 1, titled Pan-STARRS1 Source Types and Redshifts with Machine learning (PS1-STRM).

For both main data products, we use neural network architectures, trained on a compilation of public spectroscopic measurements that has been cross-matched with PS1 sources.

We quantify the parameter space coverage of our training data set, and flag extrapolation using self-organizing maps. We perform a Monte-Carlo sampling of the photometry to estimate \photoz\ uncertainty.

The final catalogue contains $2,902,054,648$ objects. On our validation data set, for non-extrapolated sources, we achieve an overall classification accuracy of $98.1\%$ for galaxies, $97.8\%$ for stars, and $96.6\%$ for quasars. 

Regarding the galaxy \photoz\ estimation, we attain an overall bias of $\left<\Delta z_{\mathrm{norm}}\right>=0.0005$, a standard deviation of $\sigma(\Delta z_{\mathrm{norm}})=0.0322$, a median absolute deviation of $\mathrm{MAD}(\Delta z_{\mathrm{norm}})=0.0161$, and an outlier fraction of $O=1.89\%$.

The catalogue will be made available as a high-level science product via
the Mikulski Archive for Space Telescopes at \url{https://doi.org/10.17909//t9-rnk7-gr88}.

\end{abstract}


\maketitle

\section{Introduction}

Optical broad-band imaging surveys have played a vital role in efforts to gather information about the Universe. 
The previous largest such survey, the Sloan Digital Sky Survey \citep[SDSS]{York2000, Blanton2017}, with a combination of imaging and spectroscopic measurements covering $\approx 14,000$ square degrees, 
provided a collection of data that has enabled decades of scientific research, resulting in over $7,700$ peer-reviewed publications\footnote{https://www.sdss.org/science/}.

The Pan-STARRS1 $3\pi$ survey\footnote{https://panstarrs.stsci.edu/}, with its Data Release 1 (PS1 $3\pi$ DR1), has become the currently largest public imaging survey with over $3$ billion unique sources occupying $\approx 30,000$ square degrees \citep{Chambers2016}. Various data products have been generated by a complex data reduction pipeline, and made publicly available \citep{Magnier2016, Flewelling2016}.
Building on these foundations, higher level science products facilitate fulfilling the scientific potential of such a vast trove of data.

When only broad-band photometry is available, the task of correctly assigning light sources to given classes of astronomical objects, e.g. stars, galaxies, or quasars, is far from trivial. 

For separating galaxies from non-galaxies, the common approach makes use of the fact that galaxies are extended sources, and defines a cut on their point spread function (PSF) aperture magnitude versus an extended aperture magnitude \citep{Scranton2002}. Difficulties arise for faint galaxies whose outer regions fade into the background, and when high seeing significantly blurs point sources. 

Quasar/star separation is more intricate, and usually relies on cuts in colour-colour diagrams \citep{Schneider2002}, but higher-redshift quasars can often only be distinguished using both optical and infrared observations \citep{Wu2010}, or
time-domain observations \citep{Schmidt2010}.

However, the choices of these traditional classification approaches --- e.g. the boundary definition, or which photometric bands and aperture types to use --- are often based on a few low-dimensional projections of a complex high-dimensional space.
On the other hand, modern machine learning methods enable us to automatically
make such choices in a data-driven way, 
utilizing the entire multi-dimensional parameter space \citep{Gao2008, Kim2015, Makhija2019}.

In the case of galaxies, a measurement of distance is essential in extracting physical and cosmological information. 
Estimating the redshift from broad-band data, i.e. photometric redshift (\photoz) estimation, has become a staple of extragalactic and cosmology research. Most imaging surveys have a corresponding \photoz\ catalogue, including the Canada-France-Hawaii Telescope
Legacy Survey \citep{Brimioulle2008}, SDSS \citep{Beck2016b}, Dark Energy Survey \citep{Hoyle2018}, and Kilo-Degree Survey \citep{Bilicki2018}.

There are two main branches of \photoz\ estimation, spectral template fitting approaches \citep{Benitez2000, Bolzonella2000, Arnouts2002, Coe2006, Ilbert2006, Brammer2008, Beck2017b}, and machine learning approaches \citep{Wadadekar2005, Csabai2007, Collister2007, Carliles2010, Gerdes2010, Brescia2014, Cavuoti2015, Beck2016b}. A number of method comparison papers have evaluated the strengths and weaknesses of each \citep{Csabai2003, Hildebrandt2010, Dahlen2013, Beck2017, Amaro2019}. Typically, machine learning approaches enable superior performance, as long as there is a sufficient number of spectroscopic redshifts available to calibrate the model.

In this paper, we present Pan-STARRS1 Source Types and Redshifts with Machine learning (PS1-STRM), the neural network source classification and \photoz\ catalogue created for PS1 $3\pi$ DR1, with a detailed description of the data processing and methodology. In Sect.~\ref{sec:dataset}, we review the data sets that supported this work. In Sect.~\ref{sec:methodology}, we describe our methods and processing steps. In Sect.~\ref{sec:results}, we present validation results and quality metrics. We summarize in Sect.~\ref{sec:conclusions}. Finally, in Appendix~\ref{sec:appendix1} and \ref{sec:appendix2} we provide details about the public database, and list some caveats.

\section{Data sets}

\label{sec:dataset}

\subsection{PS1 \texorpdfstring{$3\pi$}{3π} DR1}

\label{sec:photodata}

The first data release of PS1 $3\pi$ contains detections of $10.7$ billion unique objects over $3/4$ of the entire sky, with more than $3$ billion sources confirmed in multiple bands. 

Broad-band photometric measurements
have been taken in the PS1 $g$, $r$, $i$, $z$ and $y$ filters \citep{Tonry2012}. Refer to \citet{Chambers2016} for details about the survey, to \citet{Magnier2016} about the data processing system,
to \citet{Waters2016} about pixel processing and stacking, to \citet{Magnier2016b} about source detection and to \citet{Magnier2016c} about astrometric and photometric calibration.
Finally, a detailed description of the public database\footnote{http://mastweb.stsci.edu/ps1casjobs/} is provided in \citet{Flewelling2016}.

Reduced photometry has been produced with different methodologies to serve various scientific use cases. There is mean photometry based on single-epoch detections, stack photometry created by stacking
all observations in a given field and filter and forced mean photometry for objects detected in the stacks, but not necessarily detected in single exposures. 
All of these data products offer a selection of apertures with which fluxes and magnitudes have been extracted, including different fixed radius apertures,
point spread function (PSF) apertures, Kron apertures and seeing-matched apertures. Individual single-epoch detections were not included in DR1, but have since been published with Data Release 2.

Furthermore, for extended sources detected in the stacks, there are magnitudes available for de Vaucouleurs, Sersic, exponential and Petrosian shape fits, and additional fixed radius aperture magnitudes.

It is important to note that the catalogue is highly inhomogeneous in terms of the number and exposure time of observations (and thus depth), and in terms of the measurement quality across broad-band filters.

For the purposes of creating a uniform classification and \photoz\ catalogue for static sources in the entirety of PS1 $3\pi$, we require the deepest and most accurate photometry that is available for all sources,
and with an aperture selection that can sufficiently describe galaxies, stars, as well as quasars. The stack photometry would be the obvious candidate. 
However, as stacks are overlapping, many sources are present in multiple stacks, and accordingly in the database there are multiple rows for many unique \textit{ObjID}-s. Selecting for the best stack measurement
for every source would make processing the entire database via the public interface unfeasible in terms of execution time ($\approx 2$ years, depending on the workload of the server).

Thus, as the second deepest option, the forced mean photometry was selected for this project, specifically the \textit{ForcedMeanObject} table of the database. It contains magnitudes for 
PSF, Kron and seeing-matched apertures (\textit{FPSFMag}, \textit{FKronMag} and \textit{FApMag}, respectively), as well as $3.00 ''$, $4.63 ''$ and $7.43 ''$ fixed-radius apertures (\textit{FmeanMagR5},
\textit{FmeanMagR6} and \textit{FmeanMagR7}). Also, there is only a single forced measurement for each unique source, and it is available for every source in PS1 $3\pi$. 

We note that the Kron radius determined by the photometry pipeline is not available in either the \textit{ForcedMeanObject} or the \textit{MeanObject} tables; however, it has been published for the stack photometry.

\subsection{Combined spectroscopic sample}

\label{sec:spectrodata}

\begin{table*}
\caption{The cross-matched source counts and used quality flags of the different surveys comprising our combined spectroscopic sample.}
\label{tab:spectrosample}
\begin{center}
\begin{tabular}{l r r r r l}
	Survey & Total source count & Galaxies &  Stars & Quasars & Quality flags \\ \hline \hline
	SDSS DR14  & $3,616,323$ & $2,310,690$ & $766,251$ & $539,382$ & $\mathrm{zWarning} = \mathtt{0x00}, \, \mathtt{0x10}$ \\ 
	DEEP2 DR4  & $18,636$ & $17,143$ & $631$ &  $862$ & $\mathrm{ZQUALITY} = 4$ \\  
	VIPERS PDR-2 & $53,833$ & $51,523$ & $2,310$ & - & $\lfloor \mathrm{zflg} \rfloor \pmod{10} = 3, \, 4$\\ 
	WiggleZ  & $146,686$ & $146,647$ & $39$ &  - & $\mathrm{Q} = 4, \, 5$ \\ 
	zCOSMOS DR3  & $11,867$ & $11,125$ & $742$ & - & $\lfloor \mathrm{CC} \rfloor \pmod{10} = 3, \, 4$\\ 
	VVDS  & $6,374$ & $6,374$ & - &  - &  $\mathrm{ZFLAGS} \pmod{10} = 4$\\ \hline
	Combined  & $3,853,719$ & $2,543,502$ & $769,973$ &  $540,244$ &  \\ 
\end{tabular}
\end{center}

\end{table*}

\begin{figure*}
\begin{center}
\includegraphics[draft=false,width=\textwidth]{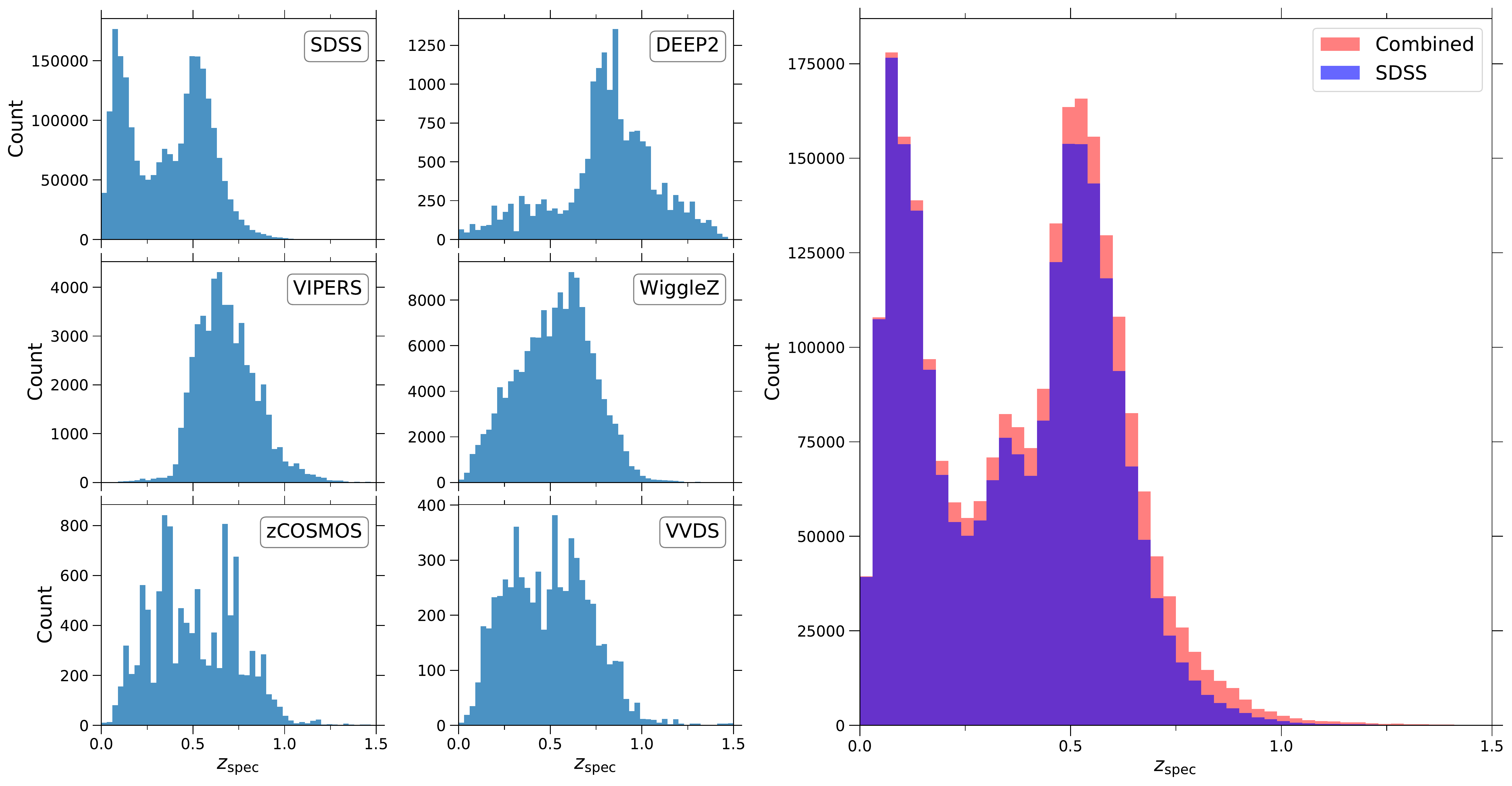}
\vspace*{-0.5cm}
\end{center}
\caption{The redshift distribution of galaxies in the spectroscopic surveys that constitute our combined spectroscopic sample. Left panel: surveys are shown individually. Right panel: the combined sample is plotted, 
as well as the only-SDSS component.}
\label{fig:spectrosample}
\end{figure*}

To create a reference data set in support of the classification and \photoz\ tasks, we collated public data from a number of spectroscopic surveys, where the type and redshift of objects
has been determined through a detailed analysis of their high-resolution spectra.

The spectroscopic surveys we included in this data set are the SDSS DR14 \citep{Bolton2012, Abolfathi2018}, 
the DEEP2 Redshift Survey DR4 \citep{Newman2013}, 
the VIMOS Public Extragalactic Redshift Survey (VIPERS) public data release 2 \citep[PDR-2,][]{Scodeggio2018},
the final data release of the WiggleZ Dark Energy Survey \citep{Drinkwater2018},
the zCOSMOS Data Release 3 \citep{Lilly2009}, 
and the final data release of the VIMOS VLT Deep Survey \citep[VVDS,][]{LeFevre2013}.

The spectroscopic sources have then been cross-matched with the PS1 $3\pi$ DR1 object catalogue (\textit{ObjectThin} table) following the Bayesian methodology detailed in Sect. 3.1 of \citet{Beck2016b}.
Similarly to that work, we only accept secure matches that are both closer than $1.5''$ and have a Bayes factor above $10,000$ \citep[Eq. 16 of][]{Budavari2008}. PS1 forced mean photometry has been extracted
for all cross-matched sources.

We limit our data set to measurements of high spectroscopic quality, specifically we only allow spectroscopic quality flags where the redshift is reported to be at least $98.2\%$ secure. A summary of the surveys,
matched source counts and quality flags are provided in Table~\ref{tab:spectrosample}, and the redshift distribution of galaxies in our combined spectroscopic sample is shown in Fig.~\ref{fig:spectrosample}.

The overwhelming fraction of spectroscopic matches in our sample has come from SDSS DR14. The other surveys diversify the sample due to differing targeting strategies, and 
contribute meaningfully to the high-redshift ($z>0.5$) coverage of galaxies, but stars and quasars only occur in them incidentally after the selections have been made.

\subsection{Dust maps}

\label{sec:dustdata}

The photometry available in the PS1 $3\pi$ DR1 catalogue has not been corrected for Galactic dust extinction. Our goal is to characterize sources both within and outside our Galaxy,
therefore the effect of both nearby and more distant Galactic dust has to be taken into account to arrive at the intrinsic properties of the sources.

To achieve this, we augmented our data set with two separate maps of dust extinction. 
The first is based on PS1 observations of Galactic stars, and tracks reddening out to a distance of $4.5 \mathrm{kpc}$ \citep{Schlafly2014}.
The second is based on Planck cosmic microwave background (CMB) observations, thus measuring overall extinction \citep{Planck2014}.

For all processed sources, the two $E(B-V)$ extinction values of the corresponding sky pixels in the two maps are obtained.

\section{Methodology}

\label{sec:methodology}

As described in Sect.~\ref{sec:photodata}, at our disposal we have photometry in PSF, Kron and seeing-matched apertures, alongside three fixed radii. 
These are computed for all $5$ PS1 broad-band photometric filters (g, r, i, z and y --- some may not be available in certain pointings), 
which in total yields $30$ measures of the multi-band flux of a light source.

Typically, the flux of stars and quasars is well-measured by a PSF aperture, as they are point sources to close approximation. 
The Kron aperture, on the other hand, is designed to capture a significant fraction of the total flux of an extended source, such as a galaxy.

For the purposes of \photoz\ estimation,
measuring the colours of a galaxy accurately is more significant than knowing the total flux in each band \citep{Benitez2000},
as the latter scales with both distance and physical size, while the former is connected to composition (e.g. stellar population, dust content) and is affected by redshift. 
The Kron radii for the five bands are determined independently in the PS1 pipeline, thus the magnitudes are in fact measured for different regions of a galaxy. 
To get accurate colours from these measurements, aperture correction would be required \citep[e.g.,][]{Iglesias2013}, but the corresponding radii are not published in the public database.

Of course, the availability of the three fixed-size apertures alleviates the need for aperture correction as long as a significant fraction of a galaxy is within one of them.
Still, the best radius for a given galaxy would need to be identified to ensure accurate photometry. Additionally, it has to be considered that mismatch in the central position of the
fixed apertures in different bands can lead to biased colours, and even with perfectly centered apertures, fixed-position, maximum-likelihood forced photometry entails non-negligible
colour bias \citep{Portillo2019}.

While the issues detailed above would be difficult to address when processing individual sources, we can rely on our extensive spectroscopic training set (see Sect.~\ref{sec:spectrodata}) 
to navigate through the $30+$ dimensional space of observables. A machine learning approach seems especially appropriate for this situation, specifically one that can handle
many dimensions, and even different types of inputs (magnitudes, magnitude errors, extinction values).

Neural networks represent very flexible non-linear models, and are particularly capable of recognizing useful patterns in multidimensional data via their automated learning process. 
Thus, the task of identifying the relevant magnitude for the situation, 
and of combining different magnitude measurements to mimic aperture correction can be done implicitly, simply by feeding the training data to a sufficiently complex network. Galactic extinction correction
can be done similarly, by providing the extinction map values at the given sky coordinates to the network.

We elected to use a neural network model for both the source classification and the \photoz\ estimation tasks. The specific implementation we chose is \textsc{Keras}\footnote{https://keras.io/}, 
which is essentially a high-level interface for the underlying \textsc{TensorFlow}\footnote{https://www.tensorflow.org/} deep learning library. The neural network steps were run massively parallel
on a commercial GTX 1070 graphics processing unit (GPU).

\subsection{Neural network configuration}

\label{sec:networkconfig}

For the classification task, we use a generic, densely connected neural network, with $3$ consecutive layers of $512$ neurons each. The rectified linear activation function was selected, as
it is non-linear, usually performs competitively, and derivatives are fast to evaluate \citep{Nair2010}.  Additionally, we use dropout of $0.2$ between all layers 
(tested between $0.0$ and $0.3$, in intervals of $0.1$), to enable the network to generalize better \citep{Srivastava2014}.

The output layer for the classifier network has $3$ neurons, with a softmax activation function \citep{Goodfellow2016}. The $3$ neurons correspond to the star, galaxy and quasar classes,
and the softmax ensures that the outputs sum to $1$, meaning that the outputs can be interpreted as an estimate of the probability of belonging to a given class.

For the \photoz\ regression task, we again chose a densely connected network, similarly with $3$ layers of $512$ neurons, and rectified linear activation. Here we do not use dropout, as it degraded performance (tested between $0.0$ and $0.2$, in intervals of $0.1$). The output layer is a single, linearly activated neuron, which provides the redshift output.

Both networks received the same $32$ inputs for each source, which included the $30$ different PS1 broad-band magnitude measures (refer to Sect.~\ref{sec:photodata} for details), 
and $2$ $E(B-V)$ extinction values from the PS1 and Planck extinction maps (see Sect.~\ref{sec:dustdata}). The values were normalized by subtracting the median and then dividing
by $1.349$ times the interquartile range within the training set (equivalent to transforming to zero mean and unit standard deviation in case of a normal distribution). This was only
done to speed up the initial training, as network weights are initialized to random samples from a standard normal distribution. Outlying values were clipped at $\pm 20$, while
missing values were set to $-20$.

For both networks, adding an additional layer, or doubling the neuron count in each layer does not noticeably improve performance, therefore the complexity of the networks was deemed sufficient
for our purposes. Interestingly, including magnitude error inputs does not improve performance in either network, but does slow down the training process, 
therefore we chose to take into account magnitude errors in a different way (see Sect.~\ref{sec:processing}).

\begin{figure*}
\begin{center}
\includegraphics[draft=false,width=\textwidth]{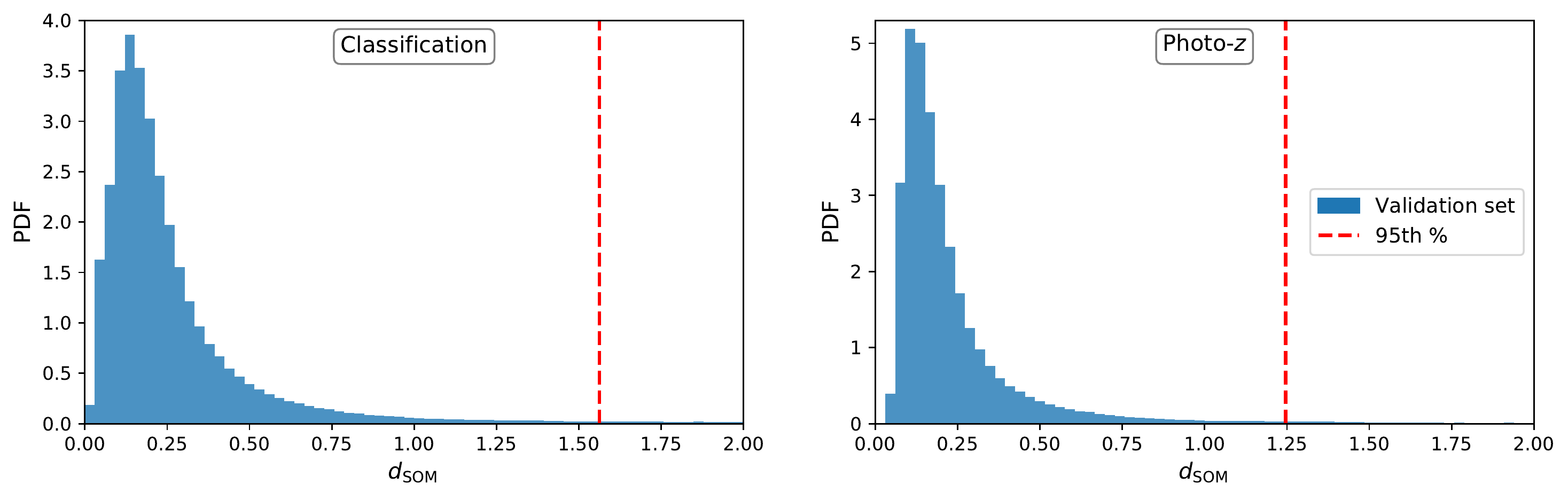}
\vspace*{-0.5cm}
\end{center}
\caption{The distribution of $d_{\mathrm{SOM}}$, the Euclidean distance (in normalized magnitude space)
from the nearest cell centre in the SOM, for validation set objects. 
Vertical dashed lines represent the cut that defines whether an object is flagged as extrapolated.
The left panel corresponds to the classification SOM and validation set, 
while the right panel shows the \photoz\ SOM and validation set.}
\label{fig:SOM}
\end{figure*}

\subsection{Training setup}

\label{sec:trainingsetup}

Our combined spectroscopic reference data set of $3,853,719$ objects (Sect.~\ref{sec:spectrodata}) has been randomly split into a training set ($80\%$ of sources) and a validation set ($20\%$).
The role of the former is to be ingested when teaching the model, i.e. setting the weights of the network, while the purpose of the latter is to monitor performance.

For the teaching algorithm, we selected backpropagation with the \textit{Adam} method, which is adaptive, and is based on the first two moments of the loss function gradient \citep{Kingma2014}. 
We use the default parameters of $\beta_1=0.9$, $\beta_2=0.999$, and a learning rate of $\mathrm{lr}=0.001$, with no learning rate decay.
The training batch size associated with these parameters is $2000$ objects.

In the case of the classifier model, the loss function we optimize is categorical cross-entropy (CCE), which is a typical choice for single-label classification (when each sample can only belong to a
single class). We teach this network for $150$ epochs, i.e. the entire training set is processed in random order $150$ times. 
For the \photoz\ regression model, the loss function is simply mean squared error, and $100$ epochs of training is sufficient for convergence.

We note that the regression model is only intended for galaxy \photoz\ estimation, and thus is only trained on galaxies. 
Specifically, it is trained on objects classified as galaxies by the classifier model (as opposed to using the spectroscopic label), mirroring the actual use case. 
Refer to Sect.~\ref{sec:resultsclass} for the definition of a successful classification.

\subsection{Self-organizing maps}

\label{sec:SOM}

While neural networks are flexible machine learning tools, there is evidence to suggest that their capacity to extrapolate into regions of the input parameter space that is not sampled by
the training set is limited \citep{Beck2017}. It is therefore important to quantify whether an object is within the effective boundaries of the training set.

In recent years, there has been a trend in \photoz\ estimation to utilize self-organizing maps (SOMs) to perform dimensionality reduction, to non-linearly project the multi-colour parameter space
into a two-dimensional grid of SOM cells in a data-driven way \citep{Masters2015, Masters2019}. We adopt this approach to quantify our training set coverage in the $30$-dimensional magnitude space.

Thus, for both the classification and the \photoz\ task, we teach a SOM on the corresponding training set, using the $30$ normalized magnitude inputs. 
The specific implementation we chose is the \textsc{SOMPY}\footnote{https://github.com/sevamoo/SOMPY} Python package. 
Using a $75 \times 150$ rectangular grid, we run $3$ epochs of rough training, and $6$ epochs of fine-tuned training, repeated for $3$ random cell starting points 
(refer to the package documentation for technical details).
Out of the $3-3$ full runs, we select the SOM with the smallest topographic error --- one SOM for the classification, and one for the \photoz.

In Fig.~\ref{fig:SOM}, we show the distribution of Euclidean distances from the nearest SOM cell centre, denoted $d_{\mathrm{SOM}}$, for both the classification and the \photoz\ validation sets.
The distributions are rather heavy-tailed; we define extrapolation as being farther from the closest SOM cell centre than the 95th percentile of $d_{\mathrm{SOM}}$ values within the validation set.

Thus, $5\%$ of objects in more outlying regions of the validation set will be considered extrapolated, which is a relatively conservative choice. 
The extrapolation limit is $d_{\mathrm{SOM}}>1.562$ and $d_{\mathrm{SOM}}>1.246$ for the classification and \photoz\ case, respectively.

\subsection{Catalogue processing}

\label{sec:processing}

This section details the steps of how every source in the PS1 $3\pi$ DR1 photometric catalog is processed, specifically every row of the \textit{ForcedMeanObjectView} database view, which
is a join between the \textit{ForcedMeanObject} and \textit{ObjectThin} tables.

For each source, first the corresponding $E(B-V)$ extinction values from the two dust maps (see Sect.~\ref{sec:dustdata})
are located, based on the $l$ and $b$ Galactic coordinates published in the database.

The $32$ neural network input fields (see Sect.~\ref{sec:networkconfig}), containing magnitude and extinction information, are normalized, then processed by the classifier model, yielding probability estimates for the galaxy, star and quasar classes.

We define a successful classification as having a predicted class probability $\mathcal{P}_\mathrm{class}>0.7$. Sources that satisfy this are accordingly flagged as galaxy, star or quasar, while sources that do not are flagged as unsure.

We note that this decision boundary is a cut on the neural network output, and does not represent a probability in the physical sense of source statistics. The choice of $0.7$ is a trade-off between limiting false positives, and missing actual class member sources. Refer to Sect.~\ref{sec:resultsclass} for an analysis of classification performance.

Using the classification SOM, we also find the nearest SOM cell for every source, and the $d_{\mathrm{SOM}}$ distance to it
(see Sect.~\ref{sec:SOM}). We flag extrapolated objects based on this distance, specifically $d_{\mathrm{SOM}}>1.562$.

The following \photoz-related steps are only performed for objects flagged as galaxies. These sources are processed by the \photoz\ neural network, yielding the base redshift estimate $z_{\mathrm{phot},0}$.

Then, similarly to the approach of \citet{Amaro2019}, we perform a Monte-Carlo sampling of the $30$-dimensional magnitude space to account for photometric errors. For each source, $100$ multivariate Gaussian random samples are drawn, with a standard deviation that matches the reported magnitude error, added as noise to the original magnitude
measurements. Each of these $100$ realizations is then processed by the \photoz\ model, generating a distribution of $100$ redshift estimates. We report the median of these values as the $z_{\mathrm{phot}}$ photometric redshift, and $1.349$ times their interquartile range as the $\Delta z_{\mathrm{phot}}$ \photoz\ error.

Finally, for each galaxy, we find the closest SOM cell in the \photoz\ SOM, and its $d_{\mathrm{SOM}}$ distance.
Galaxies having $d_{\mathrm{SOM}}>1.246$ are flagged as extrapolated.

\section{Validation results}

\label{sec:results}

In this section, we present accuracy metrics for both the neural network source classification,
and the \photoz\ regression, as computed on our validation data set (refer to Sect.~\ref{sec:methodology} for methodology
details).

\subsection{Classification}

\label{sec:resultsclass}

The classification results from our neural network are in the form of three probability-like numbers (that sum to $1$), one for each of the galaxy, star and quasar classes. 
We note that this raw neural network output has been determined to optimize a loss function in a noisy input space, and thus cannot be directly interpreted as
the physical probability of belonging to a class. To achieve a categorical assignment of sources based on the continuous output, a decision boundary has to be established.

\begin{table*}
\caption{Classification metrics for the galaxy, star and quasar classes, for different $b$ decision boundary choices: $\mathrm{P}$, the purity; $\mathrm{C}$, the completeness; and
$\mathrm{S}$, the overall success rate. The fiducial decision boundary is $b=0.7$.
The metrics were evaluated on our validation data set.
See the text for a detailed description of the metrics.}
\label{tab:classresults}
\begin{tabular}{l  l l l  l l l  l l l}
     &  & Galaxy &  &  & Star &  &  & Quasar &  \\ \hline 
	$b$ & $\mathrm{P}_\mathrm{gal}$ & $\mathrm{C}_\mathrm{gal}$ & $\mathrm{S}_\mathrm{gal}$ 
	& $\mathrm{P}_\mathrm{star}$ & $\mathrm{C}_\mathrm{star}$ & $\mathrm{S}_\mathrm{star}$ 
	& $\mathrm{P}_\mathrm{qso}$ & $\mathrm{C}_\mathrm{qso}$ & $\mathrm{S}_\mathrm{qso}$  \\ \hline \hline
0.50 & $98.03\%$ & $98.86\%$ & $97.94\%$ & $94.61\%$ & $94.11\%$ & $97.75\%$ & $90.12\%$ & $85.87\%$ & $96.70\%$ \\
0.60 & $98.30\%$ & $98.54\%$ & $97.91\%$ & $95.87\%$ & $92.62\%$ & $97.73\%$ & $92.36\%$ & $82.38\%$ & $96.57\%$ \\
\textbf{0.70} & $\mathbf{98.56\%}$ & $\mathbf{98.06\%}$ & $\mathbf{97.77\%}$ & $\mathbf{96.97\%}$ & $\mathbf{90.68\%}$ & $\mathbf{97.57\%}$ & $\mathbf{94.17\%}$ & $\mathbf{77.92\%}$ & $\mathbf{96.23\%}$ \\
0.80 & $98.82\%$ & $97.19\%$ & $97.38\%$ & $98.00\%$ & $87.88\%$ & $97.22\%$ & $95.99\%$ & $71.51\%$ & $95.59\%$ \\
0.90 & $99.13\%$ & $95.03\%$ & $96.17\%$ & $98.92\%$ & $82.93\%$ & $96.41\%$ & $97.83\%$ & $60.45\%$ & $94.27\%$ \\
0.95 & $99.34\%$ & $91.59\%$ & $94.04\%$ & $99.41\%$ & $77.64\%$ & $95.44\%$ & $98.62\%$ & $50.70\%$ & $92.99\%$ \\
0.99 & $99.75\%$ & $64.20\%$ & $76.26\%$ & $99.79\%$ & $64.55\%$ & $92.89\%$ & $99.47\%$ & $29.09\%$ & $90.04\%$ \\
\end{tabular}

\end{table*}

\begin{table*}
\caption{The same as Table~\ref{tab:classresults}, but the classification metrics were evaluated only 
on non-extrapolated sources within our validation data set.}
\label{tab:classresults2}
\begin{tabular}{l  l l l  l l l  l l l}
     &  & Galaxy &  &  & Star &  &  & Quasar &  \\  \hline 
	$b$ & $\mathrm{P}_\mathrm{gal}$ & $\mathrm{C}_\mathrm{gal}$ & $\mathrm{S}_\mathrm{gal}$ 
	& $\mathrm{P}_\mathrm{star}$ & $\mathrm{C}_\mathrm{star}$ & $\mathrm{S}_\mathrm{star}$ 
	& $\mathrm{P}_\mathrm{qso}$ & $\mathrm{C}_\mathrm{qso}$ & $\mathrm{S}_\mathrm{qso}$  \\ \hline \hline
0.50 & $98.36\%$ & $99.01\%$ & $98.25\%$ & $94.88\%$ & $95.01\%$ & $97.95\%$ & $90.85\%$ & $86.64\%$ & $97.01\%$ \\
0.60 & $98.58\%$ & $98.73\%$ & $98.22\%$ & $96.04\%$ & $93.68\%$ & $97.94\%$ & $92.92\%$ & $83.49\%$ & $96.91\%$ \\
\textbf{0.70} & $\mathbf{98.77\%}$ & $\mathbf{98.36\%}$ & $\mathbf{98.10\%}$ & $\mathbf{97.04\%}$ & $\mathbf{91.89\%}$ & $\mathbf{97.79\%}$ & $\mathbf{94.55\%}$ & $\mathbf{79.44\%}$ & $\mathbf{96.60\%}$ \\
0.80 & $98.97\%$ & $97.73\%$ & $97.82\%$ & $98.04\%$ & $89.22\%$ & $97.46\%$ & $96.20\%$ & $73.75\%$ & $96.05\%$ \\
0.90 & $99.22\%$ & $96.04\%$ & $96.88\%$ & $98.94\%$ & $84.36\%$ & $96.65\%$ & $97.89\%$ & $63.53\%$ & $94.88\%$ \\
0.95 & $99.39\%$ & $93.18\%$ & $95.10\%$ & $99.43\%$ & $79.06\%$ & $95.67\%$ & $98.65\%$ & $53.86\%$ & $93.66\%$ \\
0.99 & $99.77\%$ & $66.44\%$ & $77.67\%$ & $99.80\%$ & $65.83\%$ & $93.06\%$ & $99.49\%$ & $31.28\%$ & $90.68\%$ \\
\end{tabular}

\end{table*}

The most straightforward single-class classification requires one of the ``probabilities'' to pass a given threshold $b$, such that the assigned value for a single class will be
significantly larger than that of other classes. By choosing a $b$ decision boundary that is too high, there is a risk that sources that indeed belong to the given class
will not be classified as such, i.e. sources will be missed. On the other hand, having a decision boundary that is too low entails wrongly assigning a class to uncertain sources, 
i.e. there will be more false positives.

In our analysis, we concentrate on single-class classification outcomes (e.g., galaxy versus non-galaxy, but for all three classes), rather than listing all cross-cases.
Thus, there are four possible results, for which we introduce the following notation: 
true positive $T_1$ (e.g., galaxy classified as galaxy), true negative $T_0$ (e.g., non-galaxy classified as non-galaxy), false positive $F_1$ (e.g., non-galaxy classified as galaxy), and false negative $F_0$ (e.g., galaxy classified as non-galaxy). 

We define three metrics based on the counts of these outcomes: 
$\mathrm{P} \equiv \frac{T_1}{T_1 + F_1}$, (e.g., the fraction of true galaxies among all reported galaxies, i.e. purity), 
$\mathrm{C} \equiv \frac{T_1}{T_1 + F_0}$ (e.g., the fraction of true galaxies identified correctly, i.e. completeness), and
$\mathrm{S} \equiv \frac{T_1+T_0}{T_1 + T_0 + F_1 + F_0}$ (e.g., the overall successful classification rate of galaxies, i.e. overall success).

In Table~\ref{tab:classresults}, we report these metrics for the three classes, derived for different $b$ decision boundaries, on our validation data set. In Table~\ref{tab:classresults2}, we list the same metrics for non-extrapolated
validation set sources. Refer to Sect.~\ref{sec:SOM} for a description of the extrapolation flag.

As expected, increasing $b$ values yield a progressively higher 
purity, but a progressively lower 
completeness. 
The 
overall success rate peaks at $b=0.5$, the lowest value that excludes multiple classification,
but the change in this value is much smaller than for the other two metrics.

Thus, the relevant factor in selecting the $b$ boundary is the $\mathrm{P}-\mathrm{C}$ trade-off. 
Our fiducial choice is $b=0.7$, which puts a slightly larger emphasis on avoiding false positives, than on limiting
sources missed. Therefore sources with a probability output $\mathcal{P}_\mathrm{class}>0.7$ are classified into the given class, while sources that do not have a high enough output probability for any class are flagged as unsure.

The output catalog includes the fiducial discrete classification result, as well as the $\mathcal{P}_\mathrm{class}$ probability outputs, allowing users to make a choice based on their specific use case.

\subsection{\Photoz}

\label{sec:resultsphotoz}

Our neural network \photoz\ results include two estimates of the redshift, the base estimate $z_{\mathrm{phot},0}$ and the Monte-Carlo sampled $z_{\mathrm{phot}}$ (see Sect.~\ref{sec:processing} for details).
Additionally, we have an estimate of the \photoz\ error, $\Delta z_{\mathrm{phot}}$. For all validation set galaxies, the spectroscopic reference redshift $z_{\mathrm{spec}}$ is available for comparison.

We report our \photoz\ accuracy in terms of metrics that are standard in the literature \citep{Hildebrandt2010, Dahlen2013, Beck2017}.

The residuals are normalized such that $\Delta z_{\mathrm{norm}} = \frac{z_{\mathrm{phot}}-z_{\mathrm{spec}}}{1+z_{\mathrm{spec}}}$. Outliers are defined as $|\Delta z_{\mathrm{norm}}|>0.15$, with $O$ denoting the outlier fraction.
We compute the average bias $\left<\Delta z_{\mathrm{norm}}\right>$ and standard deviation $\sigma(\Delta z_{\mathrm{norm}})$,
only taking into account non-outliers, while we also calculate the median absolute deviation $\mathrm{MAD}(\Delta z_{\mathrm{norm}})$ for all galaxies.

In Table~\ref{tab:photozresults}, we report these metrics for both redshift estimates, 
computed on all (photometrically classified) validation set galaxies, and on non-extrapolated validation set galaxies. Refer to Sect.~\ref{sec:SOM} for a description of the extrapolation flag.

\begin{figure*}
\begin{center}
\includegraphics[draft=false,width=\textwidth]{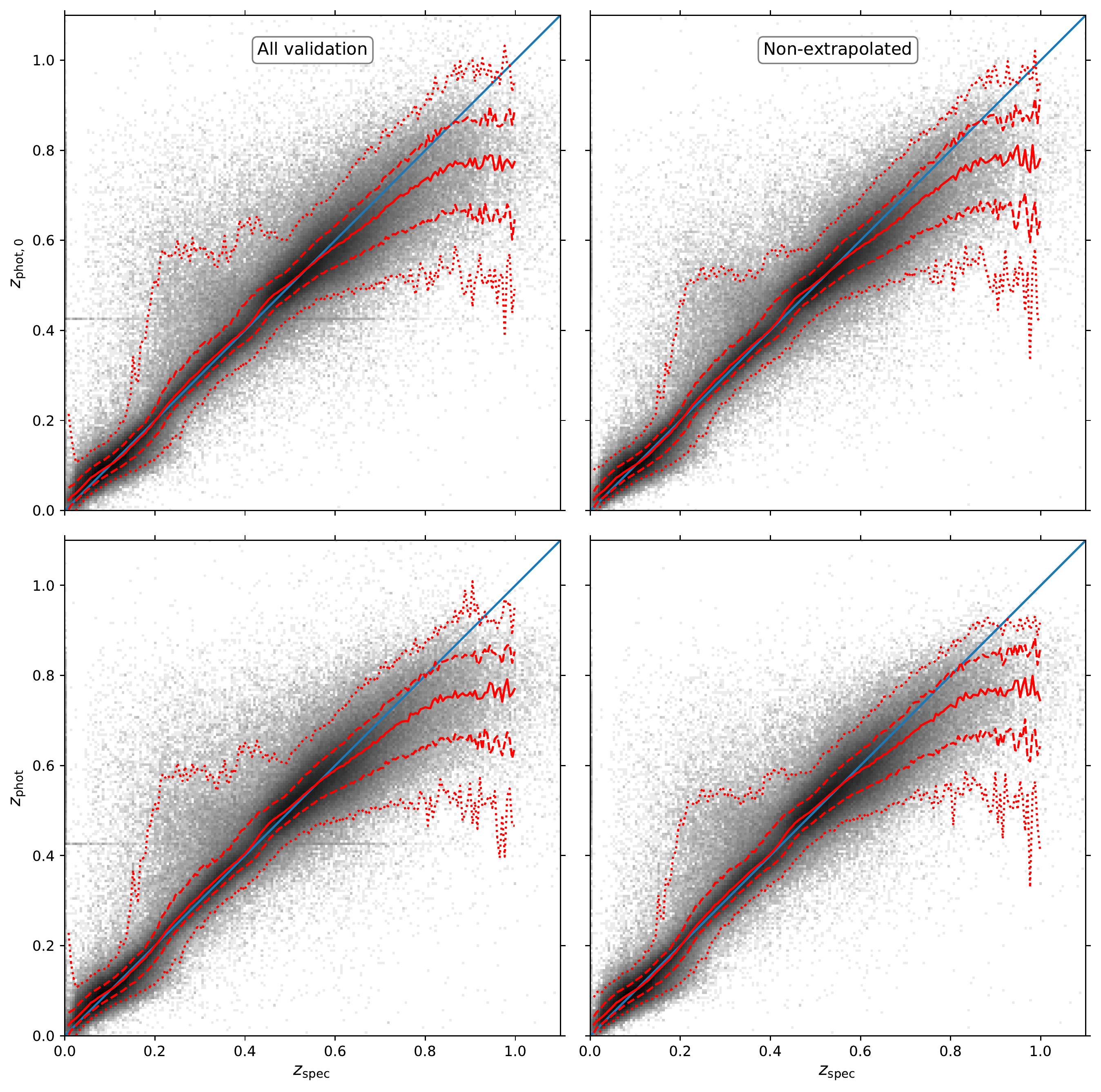}
\vspace*{-0.5cm}
\end{center}
\caption{Photometric redshift estimation results, for the base estimate $z_{\mathrm{phot},0}$ and the Monte-Carlo sampled $z_{\mathrm{phot}}$. The left column shows all validation set galaxies, while
the right column shows only non-extrapolated validation set galaxies. 
In grayscale, we plot the logarithmic density of galaxies, so that even individual objects are visible.
Solid, dashed and dotted lines show the sample median, $68\%$ confidence interval, and $95\%$ confidence interval, respectively. The main diagonal corresponds to the perfect estimation.}
\label{fig:photozresults}
\end{figure*}

Additionally, in Fig.~\ref{fig:photozresults}, we show the $z_{\mathrm{spec}} - z_{\mathrm{phot}}$
scatterplot, for the same four cases.

While the results of the two redshift estimates are very similar, the added noise of the Monte-Carlo
sampling procedure slightly degrades performance across all metrics, especially in terms of bias. Based on this, we recommend that database users select the base estimate, $z_{\mathrm{phot},0}$, optionally checking whether it is consistent with $z_{\mathrm{phot}}$.

The overall metrics did not change drastically, however, the extrapolated sources clearly introduce unwanted features on the estimation scatterplots, including
a straight line at $z_{\mathrm{phot}} \simeq 0.43$ due to objects with missing photometry, and an increased number of 
strong outliers. The photometric catalog is expected to have a much larger proportion of extrapolated sources, likely affecting the metrics more significantly. 
Thus, users are advised to limit their analysis to non-extrapolated sources.

From Fig.~\ref{fig:photozresults}, it is clear that the useful redshift range of the catalog is $z \in [0,0.6]$. Beyond that depth, there is significant negative redshift bias as the photometric accuracy
decreases, and diminishes estimation results.

\begin{table*}
\caption{Photo-$z$ accuracy metrics computed on the base and Monte-Carlo sampled redshift estimates, 
for all validation set galaxies, and for non-extrapolated validation set galaxies. See the text for more details.}
\label{tab:photozresults}
\begin{tabular}{l  l c c c c}
Data set & Estimate & $\left<\Delta z_{\mathrm{norm}}\right>$ & $\sigma(\Delta z_{\mathrm{norm}})$ & $\mathrm{MAD}(\Delta z_{\mathrm{norm}})$ & $O$  \\ \hline \hline 
All validation & $z_{\mathrm{phot},0}$  &  0.0003 & 0.0342 & 0.0169 & $2.88\%$ \\
All validation & $z_{\mathrm{phot}}$  &  0.0010 & 0.0344 & 0.0170 & $2.99\%$ \\ \hline
Non-extrapolated & $z_{\mathrm{phot},0}$  &  0.0005 & 0.0322 & 0.0161 & $1.89\%$ \\
Non-extrapolated & $z_{\mathrm{phot}}$  &  0.0013 & 0.0323 & 0.0163 & $2.00\%$ \\
\end{tabular}

\end{table*}

We evaluate the $\Delta z_{\mathrm{phot}}$ redshift error estimate only for the recommended data cut,
specifically non-extrapolated validation galaxies and the base redshift estimate $z_{\mathrm{phot},0}$.
If the residuals were normally distributed with scatter $\Delta z_{\mathrm{phot}}$ for each galaxy, the 
$\frac{z_{\mathrm{phot},0}-z_{\mathrm{spec}}}{\Delta z_{\mathrm{phot}}}$ scaled residual distribution
would match the standard normal distribution.

\begin{figure*}
\begin{center}
\includegraphics[draft=false,width=\textwidth]{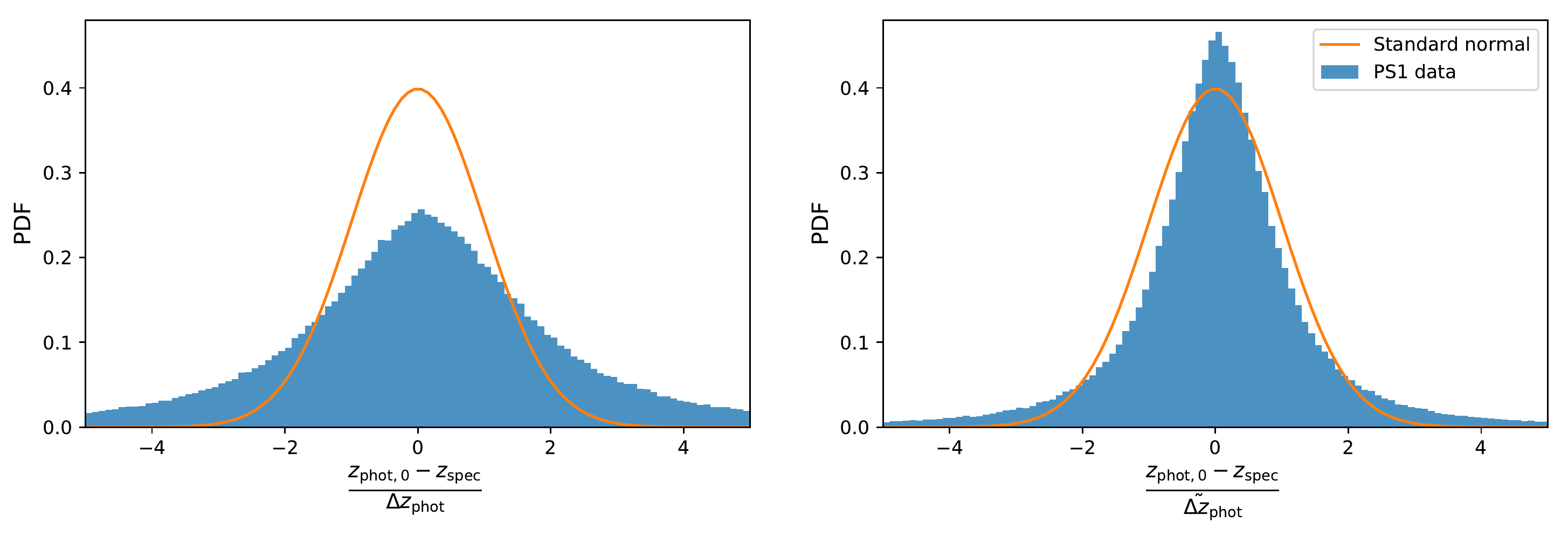}
\vspace*{-0.5cm}
\end{center}
\caption{The scaled residual distribution of the \photoz\ estimation, before and after empirical
calibration, shown alongside a standard normal distribution (solid curve). The residual distribution was computed
on non-extrapolated validation set galaxies, using the base \photoz, $z_{\mathrm{phot},0}$.}
\label{fig:photozerror}
\end{figure*}

On the left panel of Fig.~\ref{fig:photozerror}, we show the scaled residual distribution alongside a standard Gaussian. The residuals in the data are clearly larger than the estimate, indicating that
the Monte-Carlo photometry sampling did not fully capture the variance in the data. One reason
for this might be the assumption of uncorrelated errors: the photometric error of different
aperture types in the same band may not be expected to be uncorrelated.

To account for this, we can empirically calibrate the $\Delta z_{\mathrm{phot}}$ pipeline output, in essence fitting the scaled residual distribution to the standard normal distribution. 
We allow a multiplicative factor, i.e. assume $\tilde{\Delta z}_{\mathrm{phot}} \equiv A \times \Delta z_{\mathrm{phot}}$. Using $1.349$ times the interquartile range as a robust estimate of the
standard deviation, we get $A=1.986$.

The result is shown on the right panel of Fig.~\ref{fig:photozerror}. While the observed
distribution is more peaky, and has a heavier tail than a standard normal, overall $\tilde{\Delta z}_{\mathrm{phot}}$ is a reasonable estimate of the \photoz\ error within our validation set.

\section{Summary}

\label{sec:conclusions}

In this paper, we presented the methodology used in creating PS1-STRM, the neural network source classification and \photoz\ catalog
for PS1 $3\pi$ DR1, and we evaluated the quality of the data products.

We assembled a compilation of spectroscopic measurements to serve as a reference data set for machine learning algorithms.

We quantified the parameter space coverage of our training set using SOMs, and defined extrapolation boundaries based on the $d_{\mathrm{SOM}}$ distance to the nearest SOM cell.

Regarding classification, we achieve the following $\mathrm{P}$ purity, $\mathrm{C}$ completeness and $\mathrm{S}$ overall success rates, on non-extrapolated validation set sources:
for galaxies, we get $\mathrm{P}_\mathrm{gal}=98.77\%$, $\mathrm{C}_\mathrm{gal}=98.36\%$ and $\mathrm{S}_\mathrm{gal}=98.10\%$;
for stars, we get $\mathrm{P}_\mathrm{star}=97.04\%$, $\mathrm{C}_\mathrm{star}=91.89\%$ and $\mathrm{S}_\mathrm{star}=97.79\%$;
for quasars, we get $\mathrm{P}_\mathrm{qso}=94.55\%$, $\mathrm{C}_\mathrm{qso}=79.44\%$ and $\mathrm{S}_\mathrm{qso}=96.60\%$.
The trade-off between $\mathrm{P}$ and $\mathrm{C}$ can be tuned by selecting a different $b$ decision boundary.

Regarding \photoz\ estimation, on non-extrapolated validation set galaxies we achieve an overall bias of $\left<\Delta z_{\mathrm{norm}}\right>=0.0005$, a standard deviation of $\sigma(\Delta z_{\mathrm{norm}})=0.0322$, a median absolute deviation of $\mathrm{MAD}(\Delta z_{\mathrm{norm}})=0.0161$, and an outlier fraction of $O=1.89\%$.
The pipeline redshift error estimate has been empirically calibrated to the observed error distribution by setting $\tilde{\Delta z}_{\mathrm{phot}} = 1.986 \, \Delta z_{\mathrm{phot}}$.

We note that the above metrics are only applicable to non-extrapolated objects, which are expected to represent a smaller fraction of sources in the full photometric catalogue than the
spectroscopic validation set. Additionally, mismatch is expected in the distribution of other parameters, as well, possibly leading to diminished overall performance metrics on the photometric catalogue \citep{Beck2017}. For this reason, users of the database are advised to monitor the available quality-related fields of their sample,
e.g., $d_{\mathrm{SOM}}$, $\mathcal{P}_\mathrm{class}$, $\tilde{\Delta z}_{\mathrm{phot}}$ and $z_{\mathrm{phot},0}$ versus $z_{\mathrm{phot}}$.

The \photoz\ accuracy is somewhat inferior to what can be achieved with SDSS data \citep{Beck2016b}, even though the depths of the two surveys are comparable. Several notable differences might account for this. First, as discussed in Sect.~\ref{sec:methodology}, PS1 $3\pi$ DR1 does not have aperture-matched photometry, thus empirical aperture corrections are required. Second, PS1 has $y$-band observations, but does not have $u$-band. While near-infrared data is useful at high redshifts (e.g., $z>1$), when most of the light of galaxies is shifted out of the optical bands, ultraviolet instead helps constrain star-forming galaxies at low to intermediate redshifts. The latter have a more significant representation in the PS1 sample. Third, a percent-level flat-field correction has been erroneously applied with the wrong sign in PS1 $3\pi$ DR1 --- this has been fixed in DR2.

In future releases of this catalogue, from a methodological standpoint, one option to improve performance is to thoroughly optimise the hyperparameters of the neural network architecture, and report metrics for a separate, blinded validation set. This would replace our current, generic neural network setup, but the improvements from this are expected to be incremental, rather than substantial.

Including infrared observations is another option to meaningfully impact performance. Building on this work, a follow-up paper is in preparation, describing the source classification and \photoz\ catalogue created for the cross-match between WISE All-Sky \citep{Cutri2012} and PS1 $3\pi$ (Beck et al., in prep.).

The PS1-STRM catalogue\footnote{\url{http://archive.stsci.edu/hlsp/ps1-strm}} will be made publicly available upon paper acceptance as a high-level science product (HLSP) via the Mikulski Archive for Space Telescopes (MAST), managed by the Space Telescope Science Institute.

\section{Acknowledgements}

The authors thank Mark Huber for valuable discussions regarding the PS1 database.

IS and RB acknowledge support from the National Science Foundation (NSF) award 1616974, and from the National Research, Development and Innovation Office of Hungary via grant OTKA NN 129148.

The Pan-STARRS1 Surveys (PS1) and the PS1 public science archive have been made possible through contributions by the Institute for Astronomy, the University of Hawaii, the Pan-STARRS Project Office, the Max-Planck Society and its participating institutes, the Max Planck Institute for Astronomy, Heidelberg and the Max Planck Institute for Extraterrestrial Physics, Garching, The Johns Hopkins University, Durham University, the University of Edinburgh, the Queen's University Belfast, the Harvard-Smithsonian Center for Astrophysics, the Las Cumbres Observatory Global Telescope Network Incorporated, the National Central University of Taiwan, the Space Telescope Science Institute, the National Aeronautics and Space Administration under Grant No. NNX08AR22G issued through the Planetary Science Division of the NASA Science Mission Directorate, the National Science Foundation Grant No. AST-1238877, the University of Maryland, Eotvos Lorand University (ELTE), the Los Alamos National Laboratory, and the Gordon and Betty Moore Foundation.

Funding for the DEEP2 Galaxy Redshift Survey has been provided by NSF grants AST-95-09298, AST-0071048, AST-0507428, and AST-0507483 as well as NASA LTSA grant NNG04GC89G. 
The zCOSMOS survey was based on observations made with ESO Telescopes at the La Silla or Paranal Observatories under programme ID(s) 177.A-3011(E), 177.A-3011(D), 177.A-3011(H), 177.A-3011(J), 177.A-3011(B), 177.A-3011(I), 177.A-3011(C), 177.A-3011(F), 177.A-3011(G), 177.A-3011(A).
This research uses data from the VIMOS VLT Deep Survey, obtained from the VVDS database operated by Cesam, Laboratoire d'Astrophysique de Marseille, France.
This paper uses data from the VIMOS Public Extragalactic Redshift Survey (VIPERS). VIPERS has been performed using the ESO Very Large Telescope, under the "Large Programme" 182.A-0886. The participating institutions and funding agencies are listed at http://vipers.inaf.it.
The WiggleZ survey acknowledges financial support from The Australian Research Council (grants DP0772084, LX0881951 and DP1093738 directly for the WiggleZ project, and grant LE0668442 for programming support), Swinburne University of Technology, The University of Queensland, the Anglo-Australian Observatory, and The Gregg Thompson Dark Energy Travel Fund at UQ.

Funding for the Sloan Digital Sky Survey IV has been provided by the Alfred P. Sloan Foundation, the U.S. Department of Energy Office of Science, and the Participating Institutions. SDSS-IV acknowledges
support and resources from the Center for High-Performance Computing at
the University of Utah. The SDSS web site is www.sdss.org.

SDSS-IV is managed by the Astrophysical Research Consortium for the 
Participating Institutions of the SDSS Collaboration including the 
Brazilian Participation Group, the Carnegie Institution for Science, 
Carnegie Mellon University, the Chilean Participation Group, the French Participation Group, Harvard-Smithsonian Center for Astrophysics, 
Instituto de Astrof\'isica de Canarias, The Johns Hopkins University, Kavli Institute for the Physics and Mathematics of the Universe (IPMU) / 
University of Tokyo, the Korean Participation Group, Lawrence Berkeley National Laboratory, 
Leibniz Institut f\"ur Astrophysik Potsdam (AIP),  
Max-Planck-Institut f\"ur Astronomie (MPIA Heidelberg), 
Max-Planck-Institut f\"ur Astrophysik (MPA Garching), 
Max-Planck-Institut f\"ur Extraterrestrische Physik (MPE), 
National Astronomical Observatories of China, New Mexico State University, 
New York University, University of Notre Dame, 
Observat\'ario Nacional / MCTI, The Ohio State University, 
Pennsylvania State University, Shanghai Astronomical Observatory, 
United Kingdom Participation Group,
Universidad Nacional Aut\'onoma de M\'exico, University of Arizona, 
University of Colorado Boulder, University of Oxford, University of Portsmouth, 
University of Utah, University of Virginia, University of Washington, University of Wisconsin, 
Vanderbilt University, and Yale University.

The Planck dust map is based on observations obtained with Planck (http://www.esa.int/Planck), an ESA science mission with instruments and contributions directly funded by ESA Member States, NASA, and Canada.
 
\bibliographystyle{mn2e}
\bibliography{PS1_photoz}

\appendix

\section{Description of the database table}

\label{sec:appendix1}

This section gives a description of the columns of the final catalogue,
referencing the pertaining sections of the paper. For more details about fields directly reproduced from PS1 $3\pi$ DR1, refer to \citet{Flewelling2016}.

\begin{itemize}
	\item \texttt{objID} -- The main PS1 source identifier, should be used to match with other PS1 tables. Not unique.
	\item \texttt{uniquePspsOBid} -- A unique PS1 source identifier, can be used to match with \textit{MeanObject} and \textit{ObjectThin}.
	\item \texttt{raMean} -- The PS1 J2000 equatorial right ascension coordinate of the source, in degrees.
	\item \texttt{decMean} -- The PS1 J2000 equatorial declination coordinate of the source, in degrees.
	\item \texttt{l} -- The PS1 J2000 Galactic longitude coordinate of the source, in degrees.
	\item \texttt{b} -- The PS1 J2000 Galactic latitude coordinate of the source, in degrees.
	\item \texttt{class} -- The class assigned to the source, using the fiducial decision boundary $b=0.7$ (see Sect.~\ref{sec:resultsclass}). Can take the following values: ``GALAXY'', ``STAR'', ``QSO'' or ``UNSURE''.
	\item \texttt{prob\_Galaxy} -- The probability-like neural network output for the galaxy class. Corresponds to the $\mathcal{P}_\mathrm{class}$ general notation in the text. Refer to Sect.~\ref{sec:resultsclass} for more details.
	\item \texttt{prob\_Star} -- The probability-like neural network output for the star class. Corresponds to the $\mathcal{P}_\mathrm{class}$ general notation in the text. Refer to Sect.~\ref{sec:resultsclass} for more details.
	\item \texttt{prob\_QSO} -- The probability-like neural network output for the quasar class. Corresponds to the $\mathcal{P}_\mathrm{class}$ general notation in the text. Refer to Sect.~\ref{sec:resultsclass} for more details.
	\item \texttt{extrapolation\_Class} -- The extrapolation flag for the classification, $0$ if non-extrapolated, $1$ if extrapolated. Definition: $d_{\mathrm{SOM}}>1.562$ (see Sect.~\ref{sec:SOM}).
	\item \texttt{cellDistance\_Class} -- The distance to the nearest SOM cell centre in the classification SOM. Denoted $d_{\mathrm{SOM}}$ in the text (see Sect.~\ref{sec:SOM}).
	\item \texttt{cellID\_Class} -- The identifier of the nearest SOM cell in the classification SOM (see Sect.~\ref{sec:SOM}).	
	\item \texttt{z\_phot} -- The Monte-Carlo photometric redshift estimate $z_{\mathrm{phot}}$. Slightly less accurate than $z_{\mathrm{phot},0}$. Refer to Sect.~\ref{sec:processing} and \ref{sec:resultsphotoz} for more details.
	\item \texttt{z\_photErr} -- The calibrated redshift error estimate $\tilde{\Delta z}_{\mathrm{phot}} = 1.986 \, \Delta z_{\mathrm{phot}}$. Refer to Sect.~\ref{sec:processing} and \ref{sec:resultsphotoz} for more details.
	\item \texttt{z\_phot0} -- The base photometric redshift estimate $z_{\mathrm{phot},0}$. Refer to Sect.~\ref{sec:processing} and \ref{sec:resultsphotoz} for more details.
	\item \texttt{extrapolation\_Photoz} -- The extrapolation flag for the \photoz\ estimation, $0$ if non-extrapolated, $1$ if extrapolated. Definition:  $d_{\mathrm{SOM}}>1.246$ (see Sect.~\ref{sec:SOM}).
	\item \texttt{cellDistance\_Photoz} -- The distance to the nearest SOM cell centre in the \photoz\ SOM. Denoted $d_{\mathrm{SOM}}$ in the text (see Sect.~\ref{sec:SOM}).
	\item \texttt{cellID\_Photoz} -- The identifier of the nearest SOM cell in the  \photoz\ SOM (see Sect.~\ref{sec:SOM}).

\end{itemize}

\section{Caveats}

\label{sec:appendix2}

This section details several known issues concerning the catalogue.

During the construction of PS1 $3\pi$ DR1, some parts of the table \textit{ForcedMeanObject} failed to be loaded into the database. As our catalogue is based on \textit{ForcedMeanObject}, the corresponding regions are missing from our catalogue, as well.

\begin{figure}
\begin{center}
\includegraphics[draft=false,width=\columnwidth]{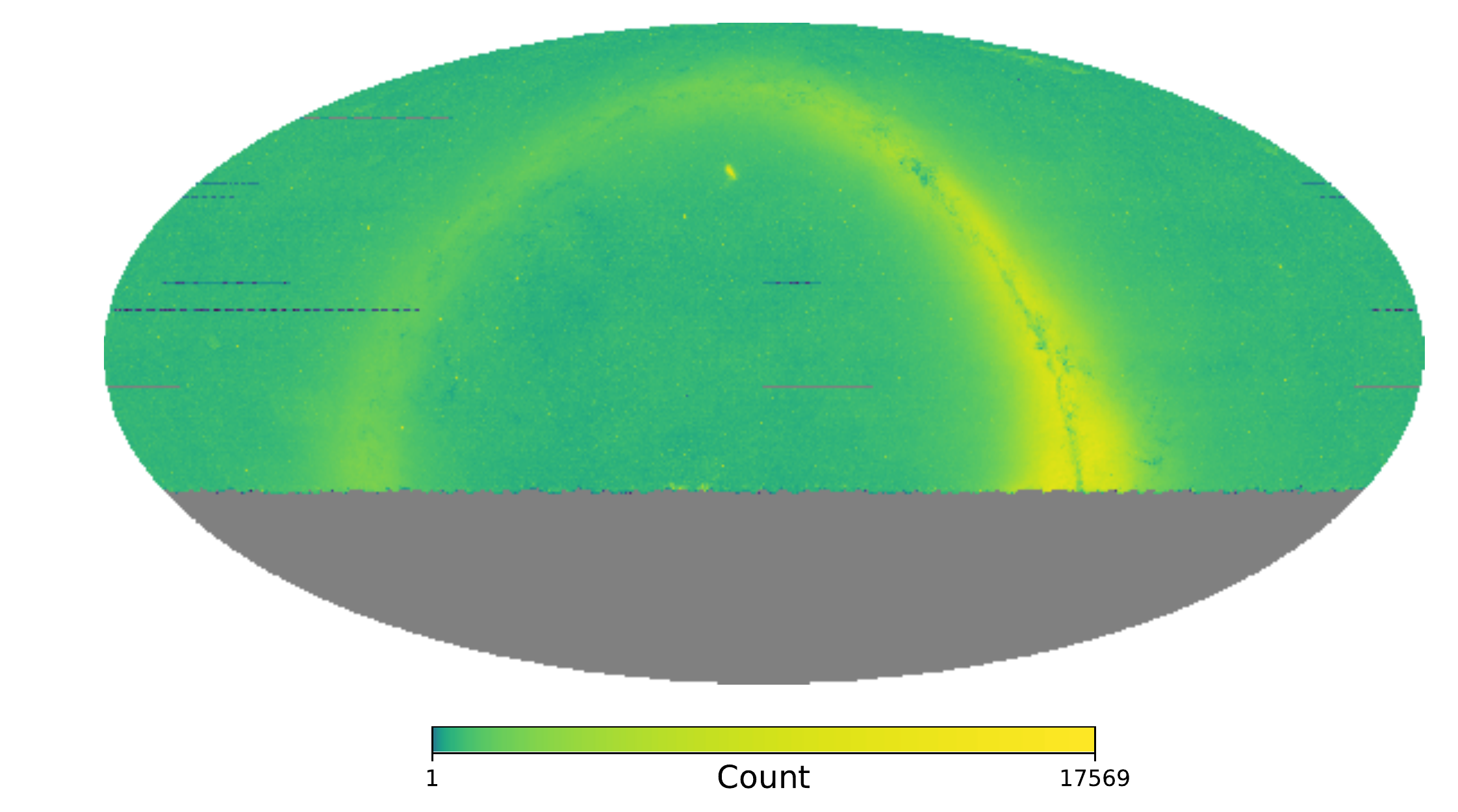}
\vspace*{-0.5cm}
\end{center}
\caption{The HEALPix source count map of our catalogue, with $\mathrm{NSIDE}=512$. The colour scale is logarithmic to illustrate a larger dynamic range. Empty cells are shown in gray.}
\label{fig:countmap}
\end{figure}

Fig.~\ref{fig:countmap} shows the HEALPix\footnote{http://healpix.sourceforge.net/} 
\citep{Gorski2005} pixelated source count map of the catalogue. Several ``lines'' of missing objects are visible, corresponding to narrow declination ranges.

Thus, users are advised to verify the availability of data when working with the following right ascension and declination ranges:
\begin{itemize}
\item $\mathrm{Dec} \in [-7.86,-7.00]$, $\mathrm{RA} \in [159.28,198.82]$
\item $\mathrm{Dec} \in [-7.86,-7.00]$, $\mathrm{RA} \in [329.45,360.00]$
\item $\mathrm{Dec} \in [8.93,9.79]$, $\mathrm{RA} \in [92.82,194.81]$
\item $\mathrm{Dec} \in [15.23,16.09]$, $\mathrm{RA} \in [131.78,168.45]$
\item $\mathrm{Dec} \in [15.23,16.09]$, $\mathrm{RA} \in [343.77,360.00]$
\item $\mathrm{Dec} \in [34.94,36.03]$, $\mathrm{RA} \in [161.57,190.22]$
\item $\mathrm{Dec} \in [38.09,39.12]$, $\mathrm{RA} \in [159.86,189.65]$
\item $\mathrm{Dec} \in [54.71,55.62]$, $\mathrm{RA} \in [118.60,186.21]$.
\end{itemize}

The choice of performing extinction correction implicitly, i.e. simply providing extinction values to the neural network, entails a drawback. Neural network models have been shown to be unreliable when extrapolating beyond the parameter coverage of the training set \citep[e.g.,][]{Beck2017}, and spectroscopy is preferentially obtained for low-extinction sources.

\begin{figure}
\begin{center}
\includegraphics[draft=false,width=\columnwidth]{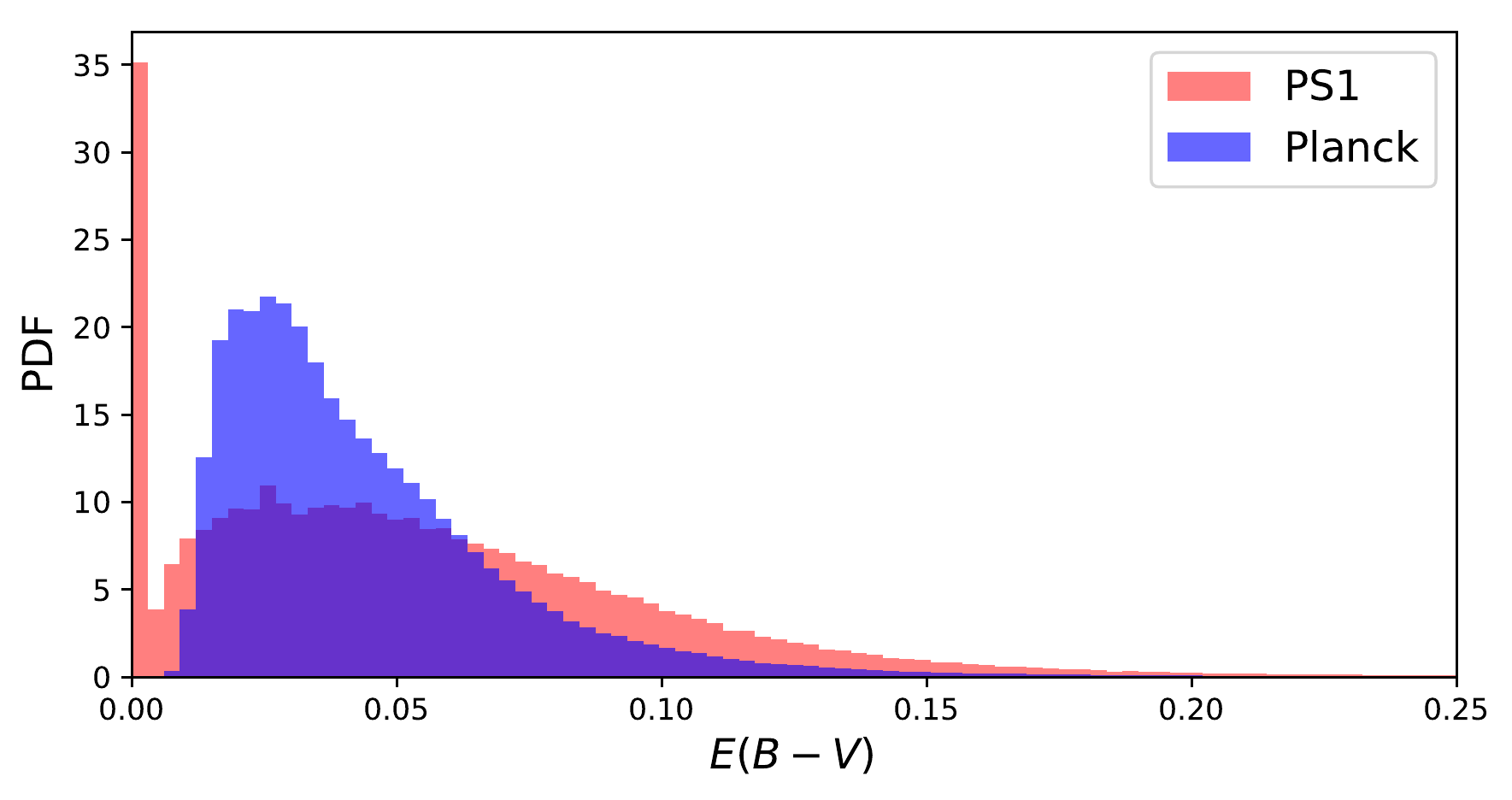}
\vspace*{-0.5cm}
\end{center}
\caption{The distribution of extinction values within the training data set, for the PS1 and Planck dust maps.}
\label{fig:extinctionhist}
\end{figure}

In Fig.~\ref{fig:extinctionhist}, we show the $E(B-V)$ distributions for our training set. Based on the sharp cut-off in the distributions, our neural network result metrics may not be expected to hold beyond $E(B-V)\simeq 0.15$ (Planck dust map) or $E(B-V)\simeq 0.2$ (PS1 dust map). Still, within our parameter coverage, explicit extinction correction performed considerably worse than the implicit in initial tests, justifying the choice.

\end{document}